\documentclass[aps,prl,twocolumn,groupedaddress]{revtex4}

\usepackage{dcolumn}
\usepackage{bm}
\usepackage{graphicx}
\usepackage{epstopdf}
\usepackage{array}
\usepackage{xcolor}
\newcolumntype{L}[1]{>{\raggedright\let\newline\\\arraybackslash\hspace{0pt}}m{#1}}
\newcolumntype{C}[1]{>{\centering\let\newline\\\arraybackslash\hspace{0pt}}m{#1}}
\newcolumntype{R}[1]{>{\raggedleft\let\newline\\\arraybackslash\hspace{0pt}}m{#1}}
\newcommand{\mgcomplex}[0]{Mg$^{2+}\cdot$(H$_2$O)$_6$}

\begin{document}

\title{Entropic modulation of divalent cation transport}

\author{Yechan Noh}
\affiliation{Department of Physics, University of Colorado Boulder, Boulder, CO 80309, USA}
\affiliation{Applied Chemicals and Materials Division, National Institute of Standards and Technology, Boulder, CO 80305, USA}
\author{Demian Riccardi}
\author{Alex Smolyanitsky}
\email[Corresponding autho r: ]{alex.smolyanitsky@nist.gov}
\affiliation{Applied Chemicals and Materials Division, National Institute of Standards and Technology, Boulder, CO 80305, USA}

\date{\today}

\begin{abstract} 
Aqueous cations permeate subnanoscale pores by crossing free energy barriers dominated by competing enthalpic contributions from transiently decreased ion-solvent and increased ion-pore electrostatic interactions. This commonly accepted view is rooted in the studies of \textit{monovalent} cation transport. Divalent cations, however, have significantly higher desolvation costs, requiring considerably larger pores to enable retention of the first hydration shell and subsequently transport. We show that this scenario gives rise to a strong enthalpy-entropy competition. Specifically, the first hydration shell is shown to undergo rotational ordering inside the pore, resulting in a tight transition state. Our results shed light on the basic mechanisms of transport barrier formation for aqueous divalent cations permeating nanoporous 2D membranes.
\end{abstract}

\maketitle
Ion transport through solvated nanoporous membranes is a fundamental process underlying complex natural and engineered systems. The thermodynamics of ion transport are broadly understood in terms of ion solvation and energetic penalties incurred during ion transitions, but the molecular-level decomposition of these phenomena remain unclear. Furthermore, thermally accessible transport depends on ion identity and valence. Here, we investigate the molecular thermodynamics of monovalent and divalent cation transport across nanoporous 2D membranes separating an aqueous ionic bath~\cite{Sahu2017, Smolyanitsky2018, fang2019highly, sahu2019optimal}. These model systems comprise two identical aqueous regions, minimally perturbed by nanoporous interfaces. Within this framework, transport barriers are informed by the electrostatic features of the pore and the membrane material; describing ion transport through such pores then reduces to a detailed understanding of all contributions to the corresponding ion-specific barriers.

The importance of nano- and sub-nanoporous 2D solids is far beyond their ability to serve as illustrative models of transport barrier formation. Recent advances in fabrication make multivacancy pores, resulting from no more than a dozen or so atomic sites ejected from the host 2D lattice, a reality~\cite{Guo2014,Liu2017,Thiruraman2020,Macha2022,Hoenig2024,Byrne2025}. In aqueous environments, permeant-specific barriers underlie unique transport properties, potentially promising to a wide range of applied areas, including molecular and ionic separation~\cite{Sun2021-jk,Violet2024}, sensing of biomolecules~\cite{Mojtabavi2019,Schneider2013} and mechanical strain~\cite{fang2019highly,fang2019mechanosensitive,sahu2019optimal,smolyanitsky2020ion,noh2024Stretch-Inactivated}, power generation~\cite{Feng2016}, and nanofluidics-based computing~\cite{noh2024memristive,noh_spiking2025,Song2025}.

\begin{figure}
\centering
\includegraphics[width=0.48\textwidth]{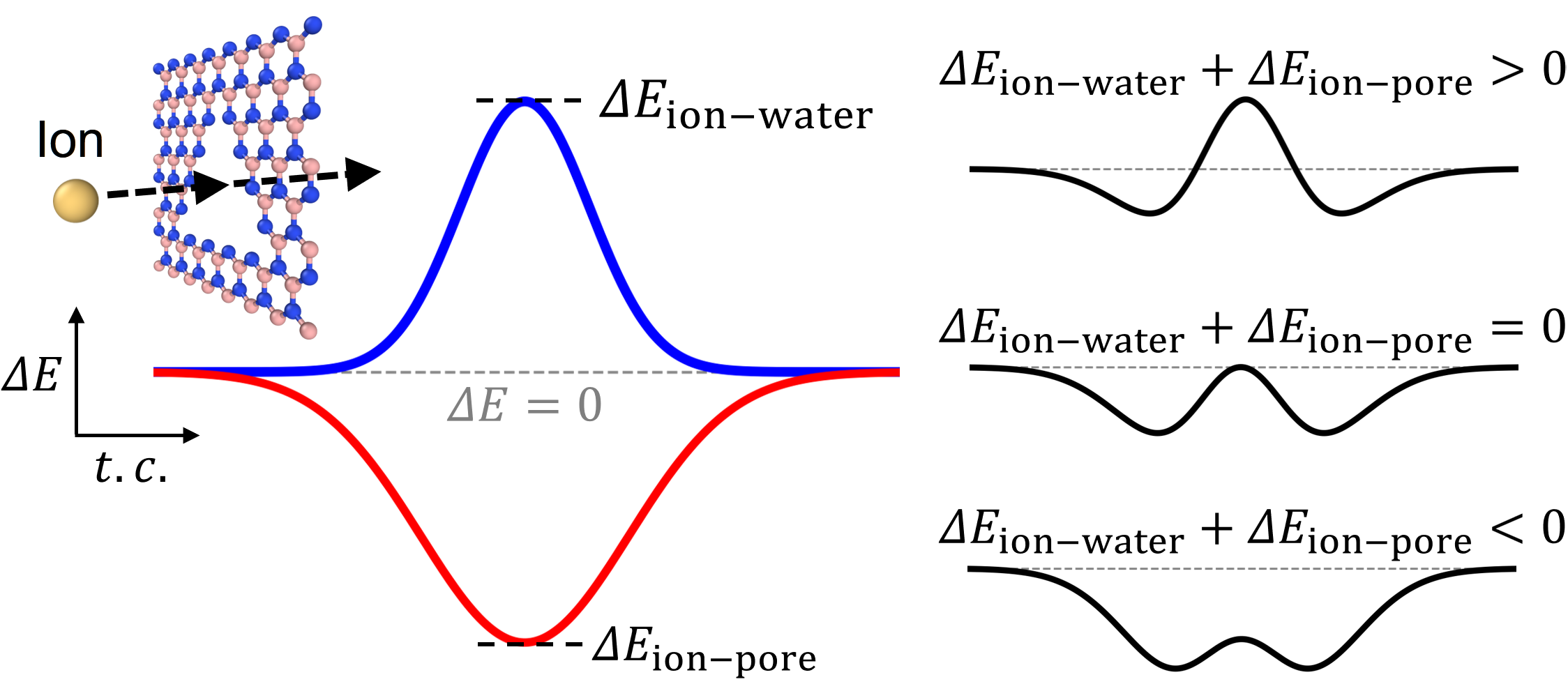}
\caption{A simplified sketch of the ion-pore and ion-water interactions as a function of the ion transport coordinate (denoted \textit{t.c.} on the left) in the direction perpendicular to the membrane plane.}
\label{fig_1}
\vspace{12pt} 
\end{figure}

Because transport of alkali salt cations is most commonly studied, theoretically and experimentally, our understanding of transport barrier formation is broadly based on the corresponding physics of monovalent cations. For subnanoscale pores with locally dipolar edges, permeation occurs one ion at a time, and the underlying mechanisms are relatively straightforward. Upon traversing the subnanoscale pore confinement, cations transiently lose a significant portion of their first hydration shell while gaining the energy of electrostatic interactions with the pore region~\cite{Sahu2017,Smolyanitsky2018,Barabash2021}. Depending on the dehydration peak height in relation to the corresponding ion-pore well depth, the overall barrier can then be attractive or repulsive, as sketched in Fig.~\ref{fig_1}. Ubiquitous in biological and artificial nanofluiodic/nanoionic systems, this apparent competition between the enthalpic ion-water and ion-pore contributions to the transport barrier was pointed out as a potentially interesting bridge with coordination chemistry~\cite{Guo2014,Smolyanitsky2018}, which describes ion interactions with entities such as crown ether molecules in aqueous environment~\cite{izatt1976calorimetric}.

Divalent cations interact with water significantly more strongly than their monovalent counterparts; for comparison, the standard enthalpies of hydration for K$^+$ and Mg$^{2+}$ are -322 kJ/mol and -1921 kJ/mol, respectively~\cite{Smith1977}. This fact makes the permeation mechanism outlined above far less probable for divalent cations under a realistic electrostatic bias. For measurable transport of divalent ions to occur, wider pores are therefore fundamentally required to allow retention of the entire first hydration shell (FHS) throughout pore traversal. 

To gain insight into transport barrier formation beyond the case of monovalent alkali cations, we investigate divalent cation permeation through nanoscale pores in monolayer hexagonal boron nitride (hBN). 
We demonstrate that upon traversal of a pore that barely permits complete retention of the FHS, a non-negligible entropic contribution to the overall transport barrier emerges, in contrast with the enthalpic view outlined above.
This contribution is shown to arise from a transient rotational immobilization of the FHS inside the pore. We also show that, depending on the membrane material, the water-water interactions can be both stabilized and destabilized by an ion entering a pore during permeation. In the main text, we focus on aqueous Mg$^{2+}$ ions, while additional results for Ca$^{2+}$ are provided in the supplementary section S3~\cite{supp}. 

\begin{figure}
\centering
\includegraphics[width=0.48\textwidth]{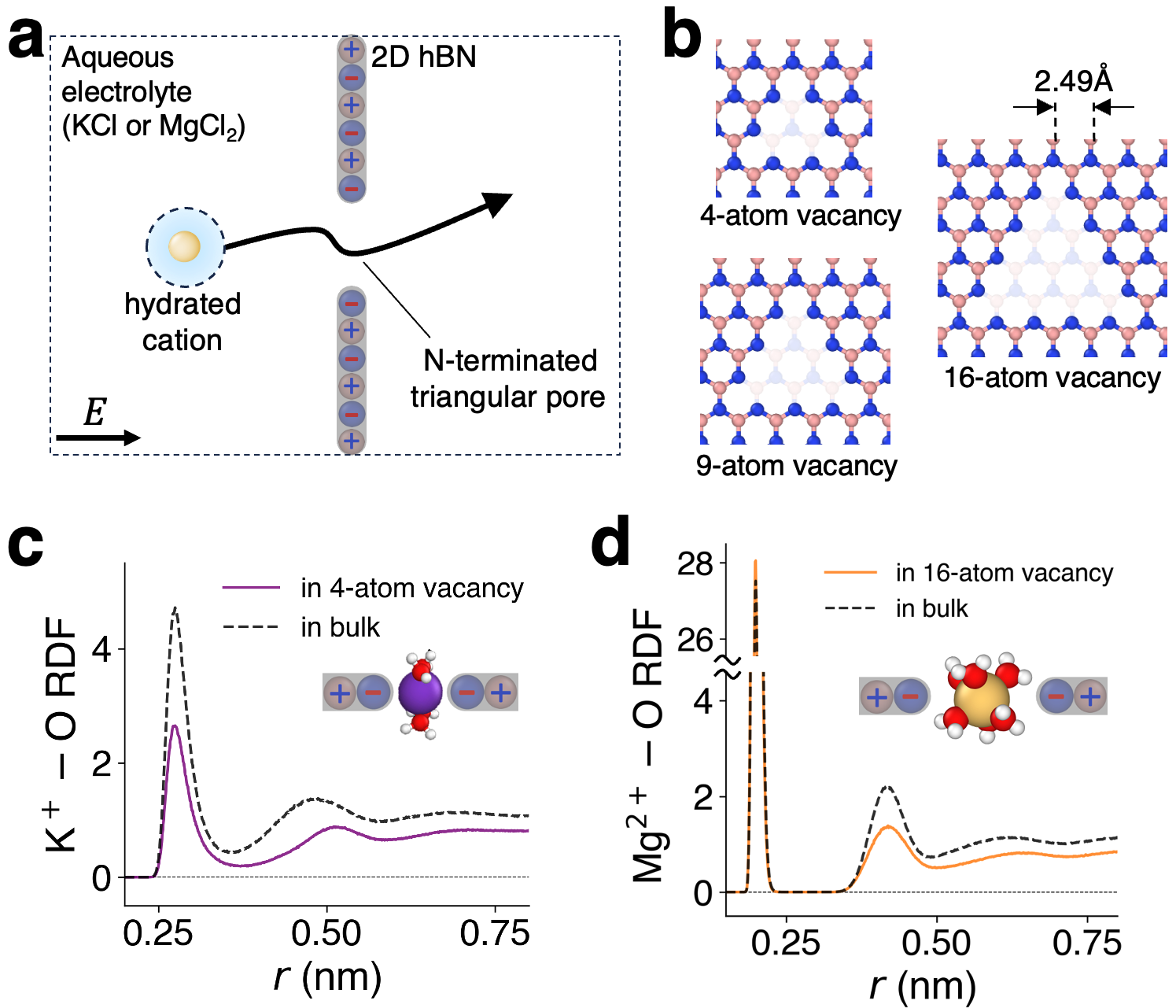}
\caption{Ion transport with transient dehydration. A general sketch of a hydrated cation permeating a nanoscale pore in aqueous hBN monolayer (a) and examples of triangular pore structures (b). Ion -- water oxygen radial distribution functions for K$^+$ cation in bulk solvent and inside a 4-atom vacancy (c) and Mg$^{2+}$ cation in bulk solvent and inside a 16-atom vacancy (d).}
\label{fig_2}
\vspace{12pt} 
\end{figure}

Our results were obtained using classical all-atom molecular dynamics (MD) simulations, performed in a rectangular 3 nm $\times$ 3 nm $\times$ 5 nm cell, periodic in $XYZ$, unless stated otherwise (\textit{e.g.}, to obtain the results of additional simulations presented in the Supplementary Material~\cite{supp}). Inside the simulation cell, a porous 2D membrane was positioned in the $XY$-plane, followed by immersion in 0.5 M of aqueous electrolyte specified later in the text. All water-dissociated salts simulated in this work were chlorides.
Simulated systems typically contained $\sim$6000 particles. Interatomic interactions were simulated within the OPLS-AA forcefield framework~\cite{jorgensen1996development}, using established parameterizations to describe hBN~\cite{govind2018ab} and MoS$_2$~\cite{sresht2017_mos2}. Additional simulations (see Supplementary Material~\cite{supp}) were carried using the GROMACS implementation of the CHARMM 36 forcefield~\cite{charmm_Bjelkmar2010}. The partial atomic charges of the edge atoms (nitrogens and S$_2$ pairs for hBN- and MoS$_2$-hosted triangular pores, respectively) were set to $2/3$ of the bulk values, resulting in charge neutrality of the pore regions in all cases. Water molecules were described according to the TIP4P model~\cite{jorgensen1983comparison}. Coulomb electrostatics was resolved using the particle-particle---particle-mesh scheme~\cite{pppm1988}. The cut-off radius for Coulomb and van der Waals interactions was 1.2 nm. A representative simulated system is sketched in Fig.~\ref{fig_2} (a) with several examples of triangular pores in hBN shown in Fig.~\ref{fig_2} (b). Atoms at the membrane perimeters were harmonically restrained to their initial positions to prevent membrane drift. Prior to production simulations, all systems underwent static energy minimization, followed by 10 ns of semi-isotropic NPT relaxation ($T=300$ K, $P=1$ bar, box $Z$-dimension adjusted, $XY$-dimensions constant). All production simulations were carried out in the NVT ensemble. Dynamic relaxation and production simulations were performed with time-steps of 1 fs and 2 fs, respectively. The potential of mean force (PMF) calculations for ion transport were performed using umbrella sampling along the transport direction ($Z$-axis) and the weighted histogram analysis method (WHAM; run with options to symmetrize the PMF)~\cite{hub2010g_wham}, as used previously~\cite{Smolyanitsky2018,fang2019highly}. All MD simulations were carried out using the GPU-accelerated GROMACS package v.2024.4~\cite{abraham2015gromacs, pall2020heterogeneous}.

\begin{figure}
\centering
\includegraphics[width=0.48\textwidth]{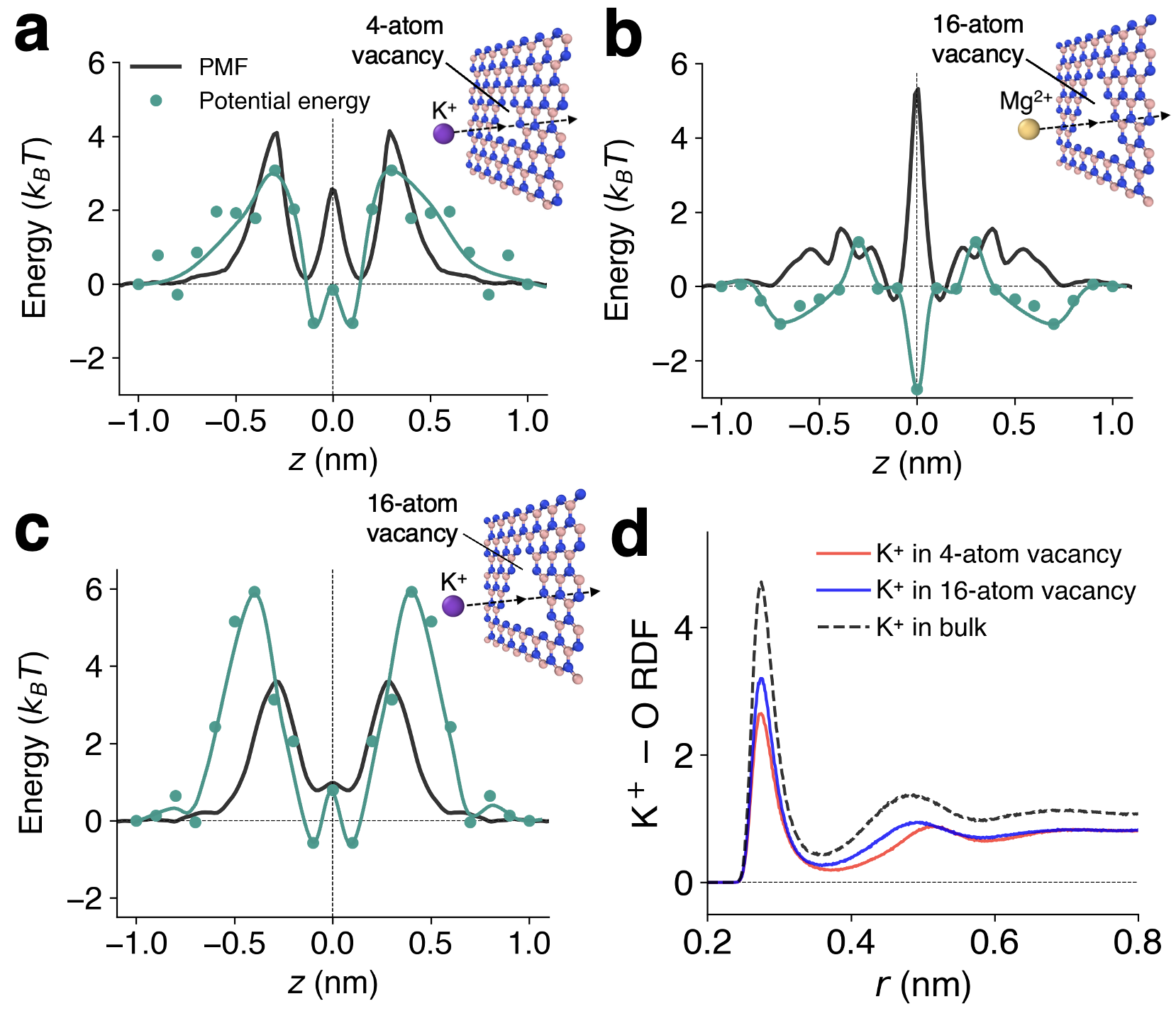}
\caption{Enthalpic contributions to the permeation energetics. Free energy profiles as a function of permeation coordinate (along Z) in the form of PMF curves, alongside the corresponding time-averaged potential energy as a function of ion's $Z$-position (solid points) for K$^+$ (a) in the 4-atom hBN vacancy, Mg$^{2+}$ in the 16-atom vacancy (b), and K$^+$ in the 16-atom vacancy (c). In (d), RDF curves for K$^+$ in selected conditions are provided. The solid green lines in (a-c) are visual guides and not numerical fits. The time-averaged potential energies were calculated on one side of the pore and mirrored in the plots above for direct comparison with the symmetric PMF; the  values are offset by the corresponding ``bulk'' value (as calculated at $Z=-1$ nm).}
\label{fig_3}
\vspace{12pt} 
\end{figure}

Consider the triangular N-terminated multivacancy pores in hBN shown in Fig.~\ref{fig_2} (b). The smallest pore permeable to aqueous K$^+$ ions is the tetravacancy~\cite{noh2024Stretch-Inactivated}, while aqueous Mg$^{2+}$ is able to permeate pores no smaller than the 16-atom vacancy pore. The effect of pore confinement on the hydration of K$^+$ is shown in Fig.~\ref{fig_2} (c). In agreement with the previously reported behavior of aqueous K$^+$ in similarly sized graphene-embedded crown ethers~\cite{Smolyanitsky2018}, the radial distribution function (RDF) of water oxygens exhibits a significant reduction, compared to bulk, in both the first and second hydration shell (SHS) around the cation located within the pore. Specifically, approximately half of the FHS is removed from a pore-confined K$^+$ ion. In contrast, from the perspective of the same RDF comparison, the larger 16-atom vacancy pore impedes divalent Mg$^{2+}$ permeation through reduced SHS interactions compared to bulk, while the FHS remains fully intact (Fig.~\ref{fig_2} (d)). However, as shown further below, the partial removal of the SHS around Mg$^{2+}$ does not reveal the source of the barrier in the PMF for the permeation of the Mg$^{2+}$ through the pore. Shown as solid curves in panels (b) and (c) of Fig.~\ref{fig_3} are the free energy profiles of Mg$^{2+}$ and K$^+$, respectively. Given the rate-setting barriers of $(5.7\pm0.5)k_BT$ for Mg$^{2+}$ and $(3.7\pm0.5)k_BT$ for K$^+$, an Arrhenius-type estimate suggests that the permeation rate for K$^+$ is between 3 and 22 times higher than that for Mg$^{2+}$. Independent simulations of ion transport in these two cases (performed at an effective transmembrane bias of 0.06 V) yielded $5.16\pm2.00$ pA and $50.36\pm3.25$ pA per pore for Mg$^{2+}$ and K$^+$, respectively. Given the charge ratio of two between Mg$^{2+}$ and K$^+$, the corresponding ratio between the permeation rates is $19.5\pm7.7$, in reasonable agreement with the estimate above. More importantly, the energetics of K$^+$ transport through the 4-atom vacancy is dominated by the potential energy of ion-pore interactions~\cite{noh2024Stretch-Inactivated}, while the presence of a repulsive peak at $Z=0$ in the Mg$^{2+}$ free energy profile (see provided visualization of simulated ionic trajectories ~\cite{supp_movie}) warrants further discussion. 

To estimate the enthalpic contributions to the free energy profiles, we carried out a series of 1-$\mu$s-long equilibrium simulations of K$^+$ and Mg$^{2+}$ harmonically restrained at various distances from the center of the corresponding pore (along the $Z$-direction). The time-averages of the potential energy (offset by a value obtained near $Z=-1$ nm, which is sufficiently far from the pore and thus close to bulk conditions) are shown as individual green points in Figs.~\ref{fig_3} (a) and (b) for K$^+$ and Mg$^{2+}$, respectively. For direct comparison of the energetics between a monovalent and divalent ion in combination with the same pore (16-atom), we show the results of the same calculations for K$^+$ in Fig.~\ref{fig_3} (c). Finally, individual contributions to the enthalpic component of the free energy are considered in the supplementary section S1~\cite{supp}.

The PMFs along the permeation coordinate agree qualitatively with the corresponding estimated enthalpy profiles for K$^+$ (Figs.~\ref{fig_3} (a) and (c)), while exhibiting a significant difference for Mg$^{2+}$ in Fig.~\ref{fig_3} (b). For K$^+$, the quantitative discrepancy between the enthalpy and the free energy in Fig.~\ref{fig_3} (a) does not appear to exceed 2$k_B T$, supporting the mostly enthalpic origin of the free energy profile, as outlined above. However, the estimated enthalpy profile for Mg$^{2+}$ in Fig.~\ref{fig_3} (b) qualitatively contrasts the PMF: the prominent repulsive barrier ($\sim6k_B T$) of the PMF near the pore mouth ($Z$=0) is in excess of $\sim8k_B T$ of the enthalpy well ($-2k_B T$). The discrepancy between the PMF and the enthalpy for Mg$^{2+}$ thus strongly suggests a large entropic contribution to the free energy barrier. 

On the timescale of membrane permeation (or order 1-2 ns), aqueous Mg$^{2+}$ can be accurately characterized as an exceptionally stable hydrated \mgcomplex\ complex. The Mg$^{2+}$ FHS lifetime in bulk water is of order microseconds~\cite{helm_lifetime1999151}, several orders of magnitude longer than it takes this ion to cross a porous 2D membrane. As described above, \mgcomplex\ permeates the pore confinement as a stable complex. We note that the transport of Mg$^{2+}\cdot$(H$_2$O)$_6$ features SHS changes that are analogous to the FHS changes for K$^+$; in fact, quantitative similarities between the corresponding dehydration costs are shown using ion-solution curves in the supplementary Figs. S1 (a) and (b).
However, we note that the six FHS waters within the \mgcomplex\ complex are generally ``configurable'' with respect to their angular locations around the Mg$^{2+}$ due to rotation of the FHS as a whole around the ion, as well as local angular deformations the FHS -- all without affecting the RDF (Fig.~\ref{fig_2} (d)). As we will show below, configurability of this three-dimensional rotor dominates the permeation barrier through a large entropic penalty that results from the reduction of rotational microstates available to the FHS as the ion approaches and traverses the pore.

We observed significant rotational immobilization of the water shell within the \mgcomplex\ complex upon crossing the pore confinement. To quantify this phenomenon, we performed a series of 20-ns equilibrium simulations, where a fully solvated Mg$^{2+}$ ion was harmonically restrained using the same PMF reaction coordinate. Throughout each simulation, we tracked the motion of each water oxygen relative to the ion and obtained the average volume available to the FHS (with the ``bulk'' value corresponding to that calculated at $Z=-1$ nm). Note that since the FHS maintains its radial structure during permeation, volume changes reflect changes in the spherical surface area statistically occupied by water oxygens. In bulk solution, the entire sphere is accessible due to the random rotational motion of the FHS around the ion (see section S2 of the Supplementary Material~\cite{supp} for details). Independently, we performed calculations of the corresponding entropic changes by tracking the spherical angles of the Mg$^{2+}$-oxygen radius-vectors within the FHS. The relative entropy is then defined as a Kullback-Leibler divergence: 

\begin{equation}
\Delta S_{KL}(N) = -Nk_B\sum\limits_{\theta,\phi} p(\theta,\phi)\ln{\left(\frac{p(\theta,\phi)}{p_0(\theta,\phi)}\right)},
\label{eq1}
\end{equation}
where $N$ is the number of FHS waters, assumed to be the same in both states subject to comparison; $p(\theta,\phi)$ is the probability density of the FHS oxygens' angular location within the two-dimensional bin cell at $(\theta,\phi)$ and $p_0(\theta,\phi)$ is the same quantity for a suitable bulk-like reference state (here, corresponding to $Z=-1$ nm). The resulting values of $T\Delta S$, alongside the bulk-normalized volume are shown in Fig.~\ref{fig_4} (a) and further details are provided in the supplementary section S2~\cite{supp}.

\begin{figure}
\centering
\includegraphics[width=0.46\textwidth]{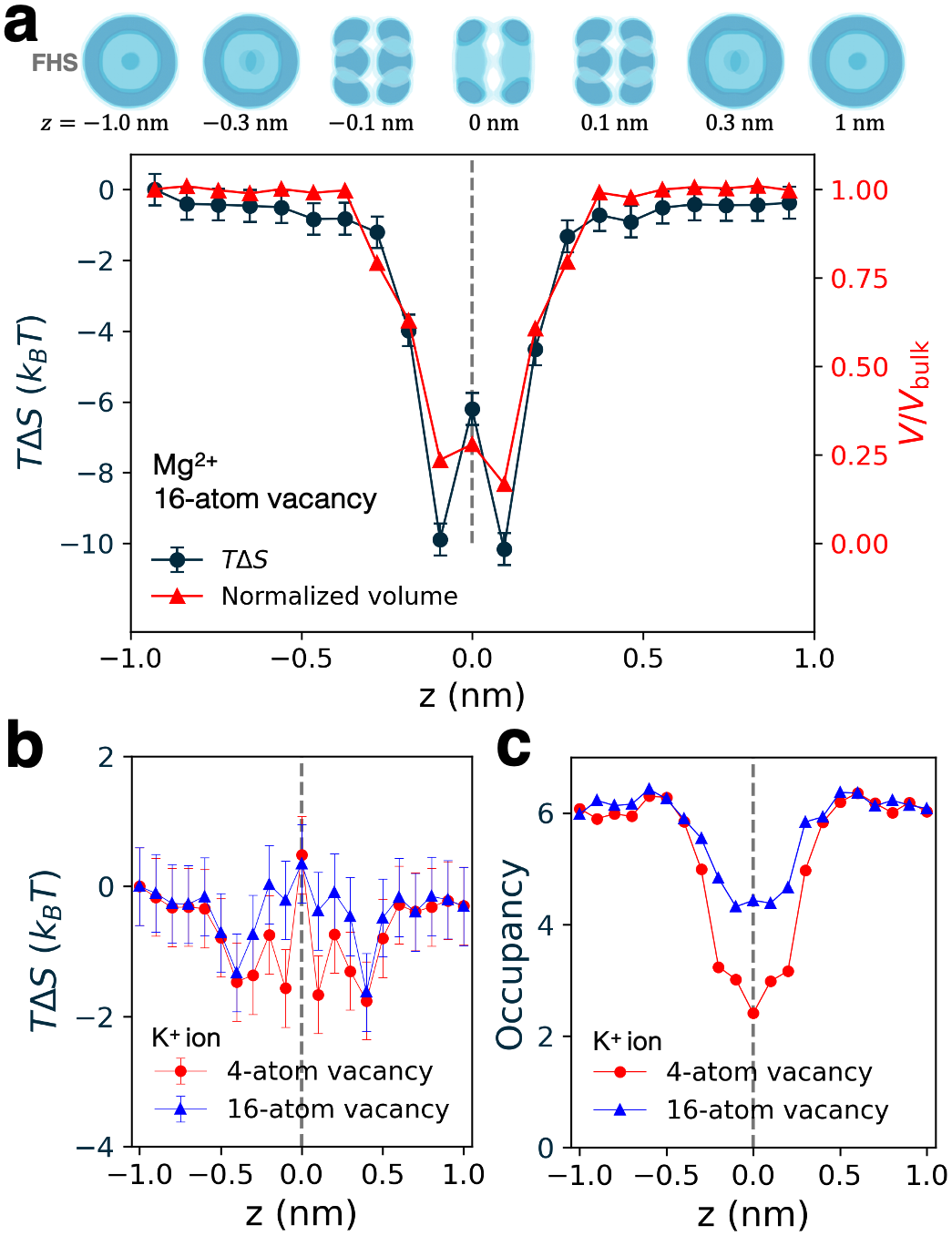}
\caption{Bulk-normalized effective volume of the FHS alongside the corresponding entropic contribution to the free energy of permeation, as a function of the $Z$-distance between the Mg$^{2+}$ ion and the 16-atom vacancy in hBN (a). As reported, $T\Delta S$ is the Kullback-Leibler entropic contribution to the free energy of permeation. Water distribution sketches around the ion are shown in blue at the top of the figure. Estimated entropy changes for the K$^+$ ion permeating the 4-atom and 16-atom pore in hBN (b), along with the corresponding FHS occupancies (c). The numerical uncertainty of entropic contributions ($\sim\frac{1}{2}k_BT$) was evaluated from divergences between physically equivalent bulk states (\textit{e.g.}, between near-bulk pairs located at $\pm Z$).}
\label{fig_4}
\vspace{12pt} 
\end{figure}

As presented, in the direct vicinity of the pore, the entropic contribution (-$T\Delta S$) to the free energy is a repulsive peak of $\sim 9k_B T$, accompanied by a highly localized 4-fold reduction in the available volume. The quantitative agreement between the magnitude of this barrier and the discrepancy observed in Fig.~\ref{fig_3} (b) clearly supports a transient loss of entropy near and inside the pore; this finding provides a clear insight into barrier formation for aqueous divalent cations. Furthermore, the entropic effects arising from rotational immobilization described above may be directly observable. For instance, Mg$^{2+}$ transport is highly mechanosensitive with pores that yield comparatively negligible mechanosensitivity with K$^+$ (see supplementary section S4~\cite{supp}).

The entropic effects described above are also present for monovalent cations, except on a more modest scale. As a result of partial FHS removal inside a pore, a purely entropic competition arises: the waters remaining in the FHS undergo a degree of angular immobilization and thus contribute a decrease in entropy. However, the waters removed from the FHS are returned to the solution, contributing a competing \textit{increase} in entropy. A more general estimate of entropy changes including these competing effects is possible; the entropy change between a reference state with FHS occupancy $N_0$ and a state with  FHS occupancy $N<N_0$ is 
\begin{equation}
\Delta S(N_0\rightarrow N) \approx \Delta S_{KL}(N) + (N-N_0)\Delta S_b,
\label{eq2}
\end{equation}
where $\Delta S_b$ is the absolute value of the per-water molecule bulk hydration entropy for a given ion. The first and second term in the sum above therefore accounts for the entropy reduction due to immobilization of what is left of the FHS inside the pore and the entropy increase associated with the removed portion of the FHS. Note that with $N=N_0$, Eq.~\ref{eq2} correctly reduces to Eq.~\ref{eq1}. The experimentally measured entropy of hydration for $K^+$ is (5.6 to 6.6)$k_B T$~\cite{Florián1999}, which, assuming six water molecules in the bulk $K^+$ FHS, yields $\Delta S_b \approx k_B$. The estimates given by Eq.~\ref{eq2} for K$^+$ in the case of 4-atom and 16-atom pore in hBN are given in Fig.~\ref{fig_4} (b). As presented, the overall entropic changes between bulk water and the pore region are modest compared with those in Fig.~\ref{fig_4} (a). Given the overall weaker levels of immobilization and considerable loss of FHS $K^+$ inside both pores (see FHS occupancy as a function of $Z$-position in Fig.~\ref{fig_4} (c)), this finding is overall expected and consistent with the PMF-enthalpy comparisons in Figs. \ref{fig_3} (b,c). For further discussion involving Ca$^{2+}$ ions, see supplementary section S3~\cite{supp}.

Typically, entropic effects play a major role in systems involving chains and polymers~\cite{Muthukumar1989,Solvik2013}, including ion-crown complexation~\cite{Michaux1982,Inoue_P29930001947}. Here, however, the effect of ``entropy on a shell'' is prominent due to the existence of an exceptionally stable ion-water complex with a high density of configurational microstates in bulk water. Given the level of rotational ordering arising from the FHS-pore electrostatic interactions (see normalized volume data in Fig.~\ref{fig_4} (a)), the resulting entropic-enthalpic interplay is strongly marked by what thermochemistry describes as a tight transition state~\cite{Gilbert1990}. 

In summary, we tuned nanopores in 2D membranes to have comparable, thermally accessible barriers for monovalent and divalent ions to carry out an in-depth analysis of transport energetics for both ion types. As expected, the nature of the permeation barrier is enthalpically driven for monovalent ions. In contrast, the permeation barrier of divalent ions is affected by an entropic cost, which increases with increasing FHS stability. The effect can be rather dramatic for Mg$^{2+}$: while this ion is enthalpically attracted to the dipolar interior of the pore, the entropic cost is shown to render the pore repulsive. Since the entropic contribution of the Mg$^{2+}$ FHS is so large, the \mgcomplex\ complex can act as a sensitive molecular switch capable of probing highly localized water-pore interactions beyond enthalpies. The entropic effect reported here appears to be experimentally testable: as shown in supplementary section S5~\cite{supp}, Mg$^{2+}$ transport is exceptionally mechanosensitive for pores that within identical strain ranges exhibit almost no such effect with monovalent cations~\cite{Noh2025diff2barrier}.

This work sheds new light on the thermodynamic driving forces underlying ion transport, providing a foundation for understanding transport phenomena beyond enthalpically driven barrier formation. Our choice of divalent cations that are known for their exceptionally stable hydration shells suggests that the findings presented here are general; other systems involving permeants with highly stable solvation shells should exhibit significant entropic contributions to barrier-limited transport. Finally, we have demonstrated once again that nanoporous atomically thin membranes in aqueous environment offer a versatile and informative framework that will continue to reveal new insights into metal ion transport.

\section{Acknowledgments}
The authors are grateful to Andrei Kazakov for illuminating discussions. Computational resources for this work were provided by the NSF's ACCESS program (Award No. PHY250014) and by the National Institute of Standards and Technology's GPU-accelerated high-performance computing facilities.

\bibliography{local}

\end{document}


\doublespacing 

\title{Supplementary Material for:

Entropic modulation of divalent cation transport}

\author{Yechan Noh}
\affiliation{Department of Physics, University of Colorado Boulder, Boulder, CO 80309, USA}
\affiliation{Applied Chemicals and Materials Division, National Institute of Standards and Technology, Boulder, CO 80305, USA}
\author{Demian Riccardi}
\author{Alex Smolyanitsky}
\affiliation{Applied Chemicals and Materials Division, National Institute of Standards and Technology, Boulder, CO 80305, USA}


\maketitle

\section{S1. Individual pairwise components of transport enthalpy}

\begin{figure*}[ht]
\centering
\vspace{12pt}
\includegraphics[width=0.8\textwidth]{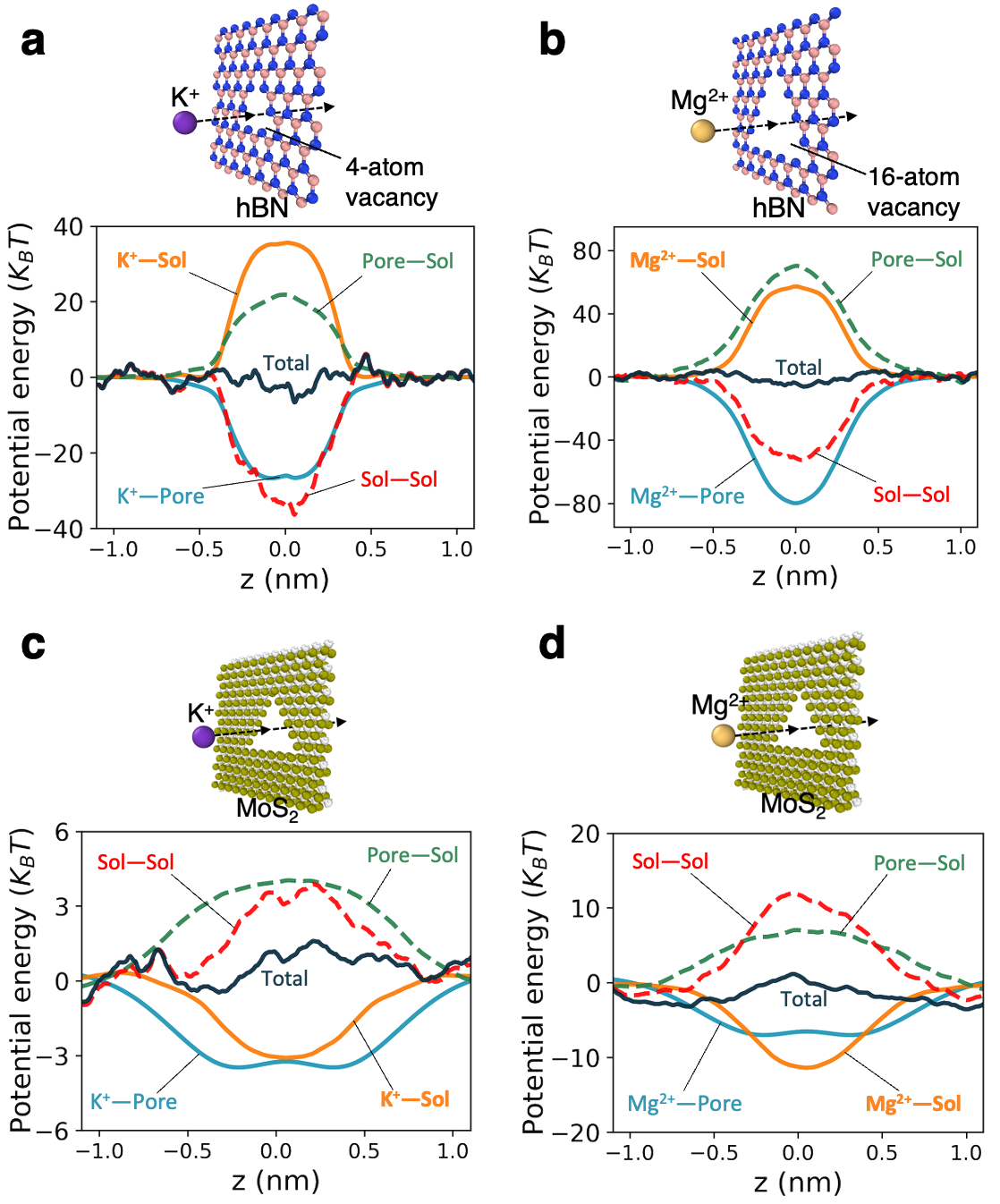}
\caption{Individual pairwise contributions to the potential energy of interactions, as obtained from pulling the K$^+$ (a) and Mg$^{2+}$ (b) along the $Z$-direction through selected pores in hBN. The same data is shown for a triangular S$_2$-terminated pore in MoS$_2$, as obtained from pulling K$^+$ (c) and Mg$^{2+}$ (d). All shown potential energies are offset by the corresponding ``bulk'' value (as calculated at $Z=-1$ nm).}
\label{fig_s1}
\vspace{18pt} 
\end{figure*}

To estimate the enthalpic contributions to ion transport, we performed 2$\mu s$-long non-equilibrium simulations of a single test ion being continuously pulled through pores in hBN along the $Z$-direction, tracking all pairwise potential energies of interaction. This included ion-solution, pore-solution, ion-pore, and solution-solution components of the total potential energy, where the solution comprises both water and dissociated ions. We extended this analysis to a MoS$_2$ nanopore to showcase qualitative differences in the enthalpic framework resulting from a change in material, highlighting the potential for material-dependent variations in ion transport; PMF and entropic characterizations of MoS$_2$ may be considered in future work. The results of these pulling simulations are presented in Fig.~\ref{fig_s1} as continuous curves.

As shown in Figs.~\ref{fig_s1} (a,b), inside the hBN pores (near $Z$=0), the pairwise components for both K$^+$ and Mg$^{2+}$ yield wells in the ion-pore (solid blue) and peaks in the ion-solvent (solid orange) component, in general agreement with the view sketched in Fig. 1 in the main text. Note that for Mg$^{2+}$ the ion-solvent peak at $Z$=0 mainly arises from partial stripping of the \textit{second} hydration shell (see complete retention of FHS in main Fig. 2 (d)). The enthalpic \textit{stabilization} of the solution-solution interactions (dashed red)  near $Z$=0 is material dependent, as shown next. 

Shown in Fig.~\ref{fig_s1} (c,d) is the same data obtained for pores of similar size and polarity in MoS$_2$. The smallest triangular pore permeable to Mg$^{2+}$ ions is obtained by removing a total of 36 atomic sites from the host MoS$_2$ lattice. All simulations used previously developed parameters~\cite{sresht2017_mos2} within the OPLS-AA forcefield framework. The atomic charges of edge sulfur atoms were set to 2/3 of their bulk values ($q_{Mo}=+0.5e$, $q_{S}=-0.25e$) to ensure an electrically neutral pore structure.
Given the size of this pore (the diameter of a circle inscribed within the pore is $\sim 1.3$ nm), neither ion in Fig.~\ref{fig_s1} (c,d) undergoes significant dehydration (including the SHS); rather, the ion-solution interactions are strengthened when the ion is inside the pore, likely due to reorganization of the SHS waters. In qualitative contrast with hBN-based pores, the solution-solution interactions in Fig.~\ref{fig_s1} (c,d) are \textit{destabilizing} for both ions. The main reason for this behavior is that an MoS$_2$ monolayer is an out-of-plane multipole, wherein a positively charged sheet formed by the partial charges of the molybdenum atoms is sandwiched between two negatively charged sheets formed by the sulfur atoms. This causes structuring/stabilization of the solution at the membrane-water interface that is destabilized when a cation enters the pore. In contrast, hBN lacks this sandwich multipole and thus a cation inside the pore  \textit{stabilizes} the water molecules within a suitable spatial bound, which results in a solution-solution energy well centered at $Z=0$. 

The observed changes in solvent behavior during ion transport between hBN and MoS$_2$ suggest that varying the material properties can significantly alter the characteristics of the transport barrier. Specifically, our findings indicate that the enthalpic competition between ion-pore and ion-solvent interactions is material-dependent, and that this dependence can be leveraged to tune the barrier properties. This insight has important implications for the biomimetic design of nanopores, where careful selection of materials and pore architectures can be used to optimize ion transport properties. In addition, these findings may help explain the highly repulsive barriers observed in smaller  MoS$_2$-based nanopores~\cite{fang2019mechanosensitive}, which require considerably larger pores than their hBN-based counterparts to enable permeation of the same ionic species.

\section{S2. Angular distributions}
\begin{figure*}[ht]
\centering
\vspace{12pt}
\includegraphics[width=0.98\textwidth]{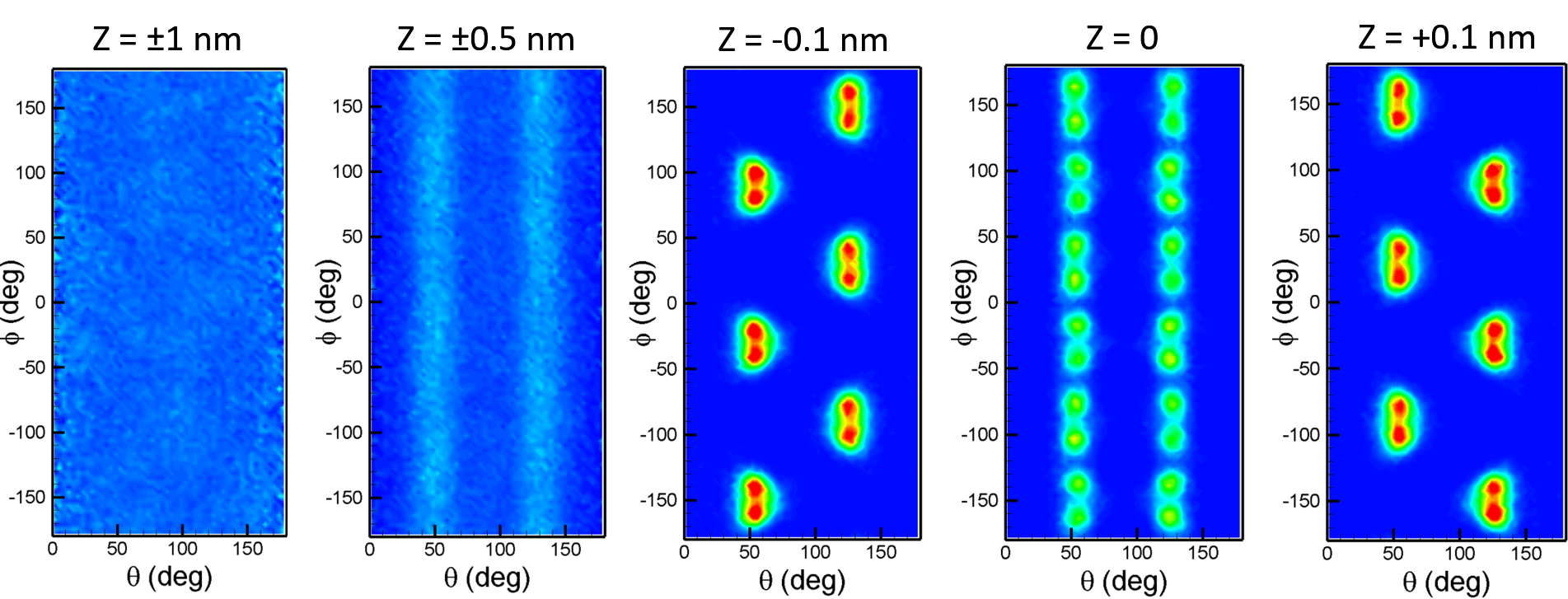}
\caption{2D histograms of the polar angles $(\theta,\phi)$ of FHS water oxygens with Mg$^{2+}$ ion as the origin in polar coordinates, obtained for various $Z$-distances from the pore.}
\label{fig_s2}
\vspace{18pt} 
\end{figure*}

We show a set of 2D histograms showing FHS immobilization when a Mg$^{2+}$ ion is inside the 16-atom pore in hBN in Fig.~\ref{fig_s2} to support the observations drawn in the main text from Fig. 4 (a). A $50 \times 100$ grid was used for the angular ranges $\theta \in \left[0,\pi\right]$, $\phi \in \left[-\pi,\pi\right]$, corresponding to $\Delta\theta = \Delta\phi=\pi/50$. The partial occupancy data used in the entropy estimation was calculated as $p(\theta,\phi)=h(\theta,\phi)\Delta A$, where $h(\theta,\phi)$ is the histogram element in Fig.~\ref{fig_s2} and $\Delta A = \sin(\theta)\Delta\theta\Delta\phi$ is the corresponding area of a spherical surface element. The partial occupancy data was normalized to yield a total of six water molecules in the FHS (calculated independently). 

As shown in Fig.~\ref{fig_s2}, the bulk-like region ($Z=\pm1$ nm) corresponds to a nearly equiprobable angular distribution throughout the entire sphere. When the ion enters the region within two solvent layers of the membrane ($Z=\pm0.5$ nm), immobilization in terms of the $\theta$-angle begins to occur. The distribution exhibits further significant immobilization in terms of both polar angles as the ion enters the pore ($|Z| \le \ 0.1$ nm). 

\section{S3. Extension to C\lowercase{a}$^{2+}$ and additional simulations of M\lowercase{g}$^{2+}$ using CHARMM 36 forcefield}
\begin{figure*}[ht]
\centering
\vspace{12pt}
\includegraphics[width=0.75\textwidth]{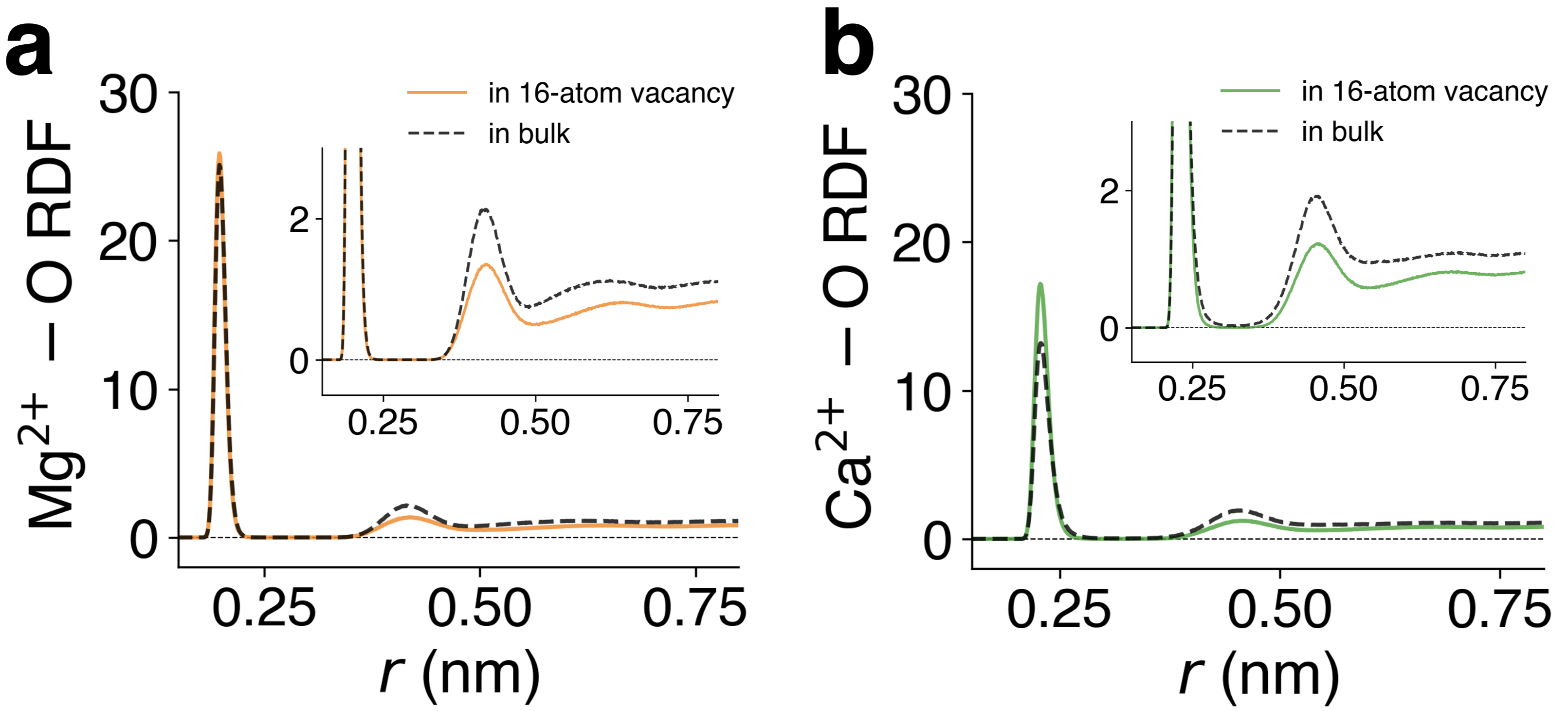}
\caption{Ion - water oxygen radial distributions for Mg$^{2+}$ (a) and Ca$^{2+}$ (b) obtained using the CHARMM 36 forcefield.}
\label{fig_s3}
\vspace{18pt} 
\end{figure*}

We extended our study to include Ca$^{2+}$ using the CHARMM 36 forcefield; for direct comparisons to the main text we also simulate Mg$^{2+}$ with CHARMM 36. CHARMM 36 simulations for Mg$^{2+}$ yield quantitative agreement with the OPLS-AA forcefield data presented in the main text. The ion-water oxygen RDFs for the ion in bulk water, as well as inside the pore (analogous to Fig. 2 (d)) are shown in Fig.~\ref{fig_s3} for Mg$^{2+}$ and Ca$^{2+}$ in panel (a) and (b), respectively. The results for Mg$^{2+}$ are nearly identical to those presented in the main text. In contrast to the static FHS of Mg$^{2+}$, the more fluidic FHS of Ca$^{2+}$ undergoes a minor \textit{gain} in hydration, corresponding to an occupancy increase from six in the bulk to approximately seven inside the pore. The associated entropic contributions to ion-pore permeation of Ca$^{2+}$ will be discussed below.  

\begin{figure*}[ht]
\centering
\vspace{12pt}
\includegraphics[width=0.98\textwidth]{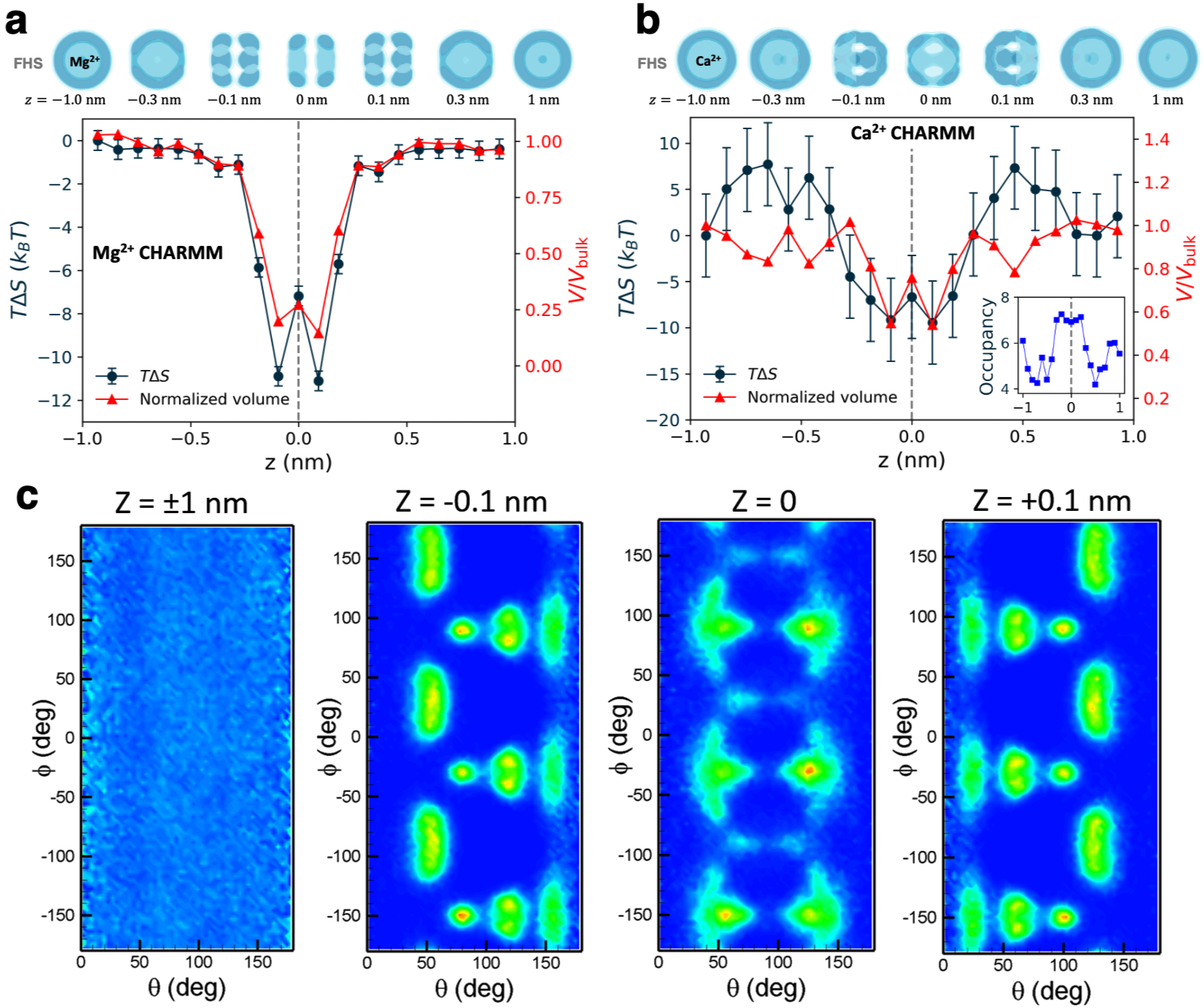}
\caption{Bulk-normalized effective volume of the FHS and the corresponding entropy relative to bulk, plotted as a function of the $Z$-distance between the ion and the 16-atom vacancy in hBN for Mg$^{2+}$ (a) and Ca$^{2+}$ (b) simulated using the CHARMM 36 forcefield. The inset in (b) is the corresponding FHS occupancy as a function of $Z$. 2D histograms of the polar angles $(\theta,\phi)$ of FHS water oxygens around the Ca$^{2+}$ ion at various $Z$-distances from the pore (c).}
\label{fig_s4}
\vspace{18pt} 
\end{figure*}

Shown in Fig.~\ref{fig_s4} are the estimates of the entropic quantities for Mg$^{2+}$ (a) and Ca$^{2+}$ (b) simulated using the CHARMM 36 forcefield paired with the TIP4P water model. The results for Mg$^{2+}$ with CHARMM 36 are nearly identical to those shown in Fig. 4 of the main text. Because the FHS occupancy of Ca$^{2+}$ increases at the pore compared to bulk (see the inset in Fig.~\ref{fig_s4} (b)), we developed a more general form of Eq. 2 in the main text that can estimate the change in entropy relative to the bulk whether $N$ is larger or smaller than $N_0$ at the pore:
\begin{equation}
\Delta S(N_0\rightarrow N) \approx \Delta S_{KL}(\min{(N,N_0)}) + (N-N_0)\Delta S_b,
\label{eq_s1}
\end{equation}
where the second term is negative when $N<N_0$ (partial loss of FHS accompanied by entropy increase caused by release of waters into bulk solvent) and positive when $N<N_0$ (partial hydration increase relative to bulk and thus a decrease in entropy). For Ca$^{2+}$, the absolute value of bulk hydration entropy is $\sim 26 k_B T$ ~\cite{Florián1999} and thus with six waters in bulk FHS, we estimate $\Delta S_b \sim 4.5 k_B$. Thus, $\Delta S_b$ for Ca$^{2+}$is significantly higher than the order $k_B T$ for K$^+$ used in the main text, but this difference is not unreasonable; the $\Delta S_b$ is around 3 to 4 times as large for Li$^+$ as K$^+$~\cite{Florián1999} assuming 4 and 6 FSH waters, repectively. The second term in Eq.~\ref{eq_s1} carries a considerable uncertainty associated with the occupancy variability in thermodynamically equivalent states (\textit{i.e.}, at $\pm Z$). We estimate the total uncertainty ($u_S$) to be $\sim 4.5 k_B T$ using the following equation, 
\begin{equation}
u_S = \epsilon_{KL} + T\Delta S_b \epsilon_{N},
\label{eq_s2}
\end{equation}
where $\epsilon_{KL}\sim 0.5 k_B T$ is estimated uncertainty of our Kullback-Leibler calculations and $\epsilon_{N} \sim 0.9$ is the estimated occupancy uncertainty for Ca$^{2+}$. 
The uncertainty of $4.5 k_BT$ is applied in Fig.~\ref{fig_s4} (b).

As presented, the entropic variation estimated using Eq.~\ref{eq_s1} is overall \textit{oscillatory}, caused by the dehydration and overhydration at $Z\sim \pm 0.5$ nm and $Z\approx 0$, respectively. This causes two entropically loose transition states on either side of the pore and a tight transition state near $Z=0$. This finding is supported in our comparison of the PMF and the enthalpic component of the free energy of permeation (direct analog of Figs. 3 (a, b) in the main text) for the Ca$^{2+}$ cation.

\begin{figure*}[ht]
\centering
\vspace{12pt}
\includegraphics[width=0.55\textwidth]{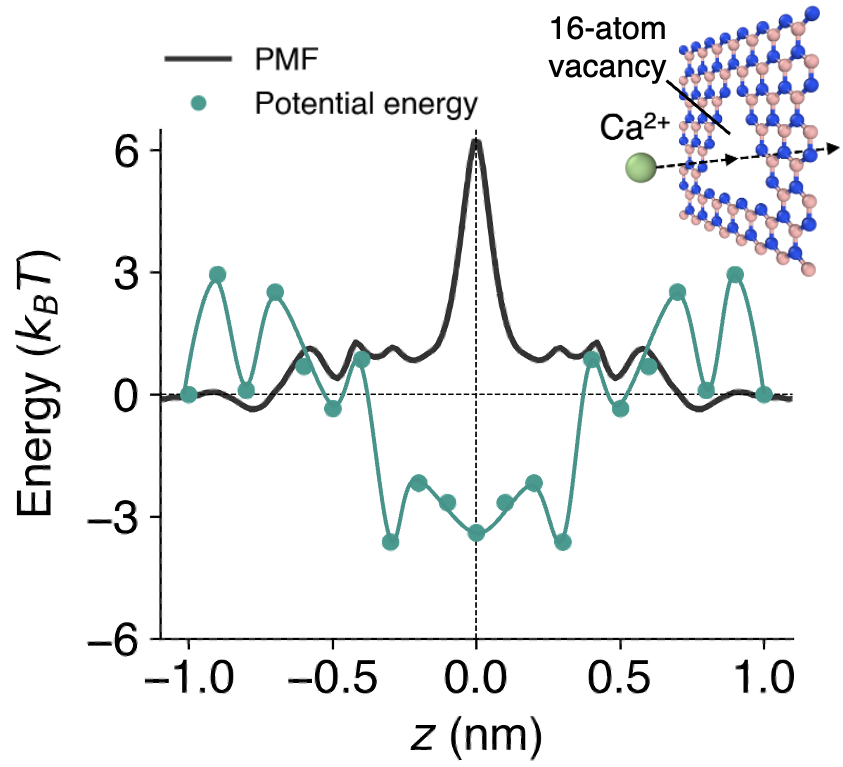}
\caption{Comparison between the enthalpic contribution to the free energy of permeation and the free energy of permeation in the form of a PMF profile for a Ca$^{2+}$ cation permeating a 16-atom pore in hBN under the CHARMM 36 forcefield. }
\label{fig_s5}
\vspace{18pt} 
\end{figure*}

Transport of Mg$^{2+}$ can therefore be viewed as a special case, in which the exceptional stability of the FHS results in $N=N_0$ throughout the permeation process. We also note that the Ca$^{2+}$ FHS immobilization in the pore region relative to bulk (the first term in Eq.~\ref{eq_s1}) is weaker than that for Mg$^{2+}$. More specifically, we calculated $-T\Delta S_{KL} \sim 6k_B T$ at $Z = \pm 0.1 $ nm and $\sim 4k_B T$ at $Z=0$, corresponding to roughly half of those observed for Mg$^{2+}$. The corresponding volume reduction is also significantly weaker, as shown in Fig.~\ref{fig_s4} (b). Finally, the angle histograms in Fig.~\ref{fig_s4} (c) confirm that Ca$^{2+}$ FHS is overall more rotationally mobile in the vicinity of the pore; for comparison, consider the histograms for Mg$^{2+}$ in Fig.~\ref{fig_s2}.

\section{S4. Sensitivity to pore size variation at sub-angstrom scale}
\begin{figure*}[ht]
\centering
\vspace{12pt}
\includegraphics[width=0.98\textwidth]{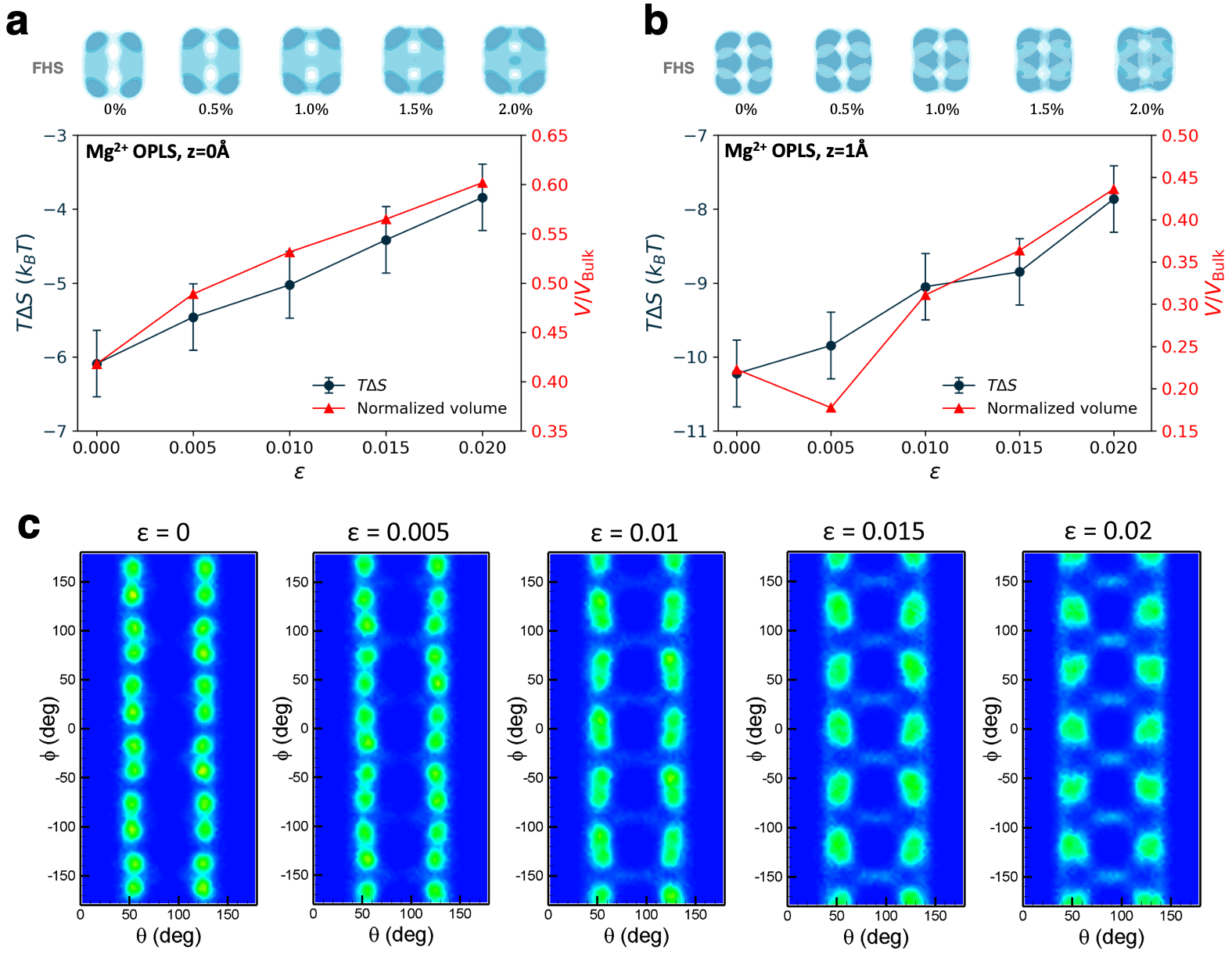}
\caption{Bulk-normalized effective volume of the FHS and Kullback-Leibler entropy change between ion in bulk ($Z = -1$ nm) and $Z=0$ (a) and $Z=0.1$ nm (b), as a function of isotropic in-plane strain. Angle histograms for various strains at $Z=0$ (c). }
\label{fig_s6}
\end{figure*}

Modulation of the enthalpic-entropic competition for a Mg$^{2+}$ ion inside a 16-atom vacancy in hBN is possible through varying the FHS-pore potential energy of interactions. The latter is achieved through applying small isotropic $XY$-strain to the pore-carrying membrane. The resulting entropy-related quantities as a function of strain are shown in Fig.~\ref{fig_s6}. As presented, both the $T \Delta S$ and the normalized volume are sensitively modulated by strain. In particular, at both $Z=0$ (Fig.~\ref{fig_s6} (a)) and $Z=0.1$ nm (Fig.~\ref{fig_s6} (b)), $-T \Delta S$ modulation reaches $\sim 2k_B T$ at 2\% strain. Gradual reduction in FHS immobilization with increasing strain is further confirmed in Fig.~\ref{fig_s6} (c), where we show the angle histograms at $Z=0$ for various strain values. 

\begin{figure*}[ht]
\centering
\includegraphics[width=0.55\textwidth]{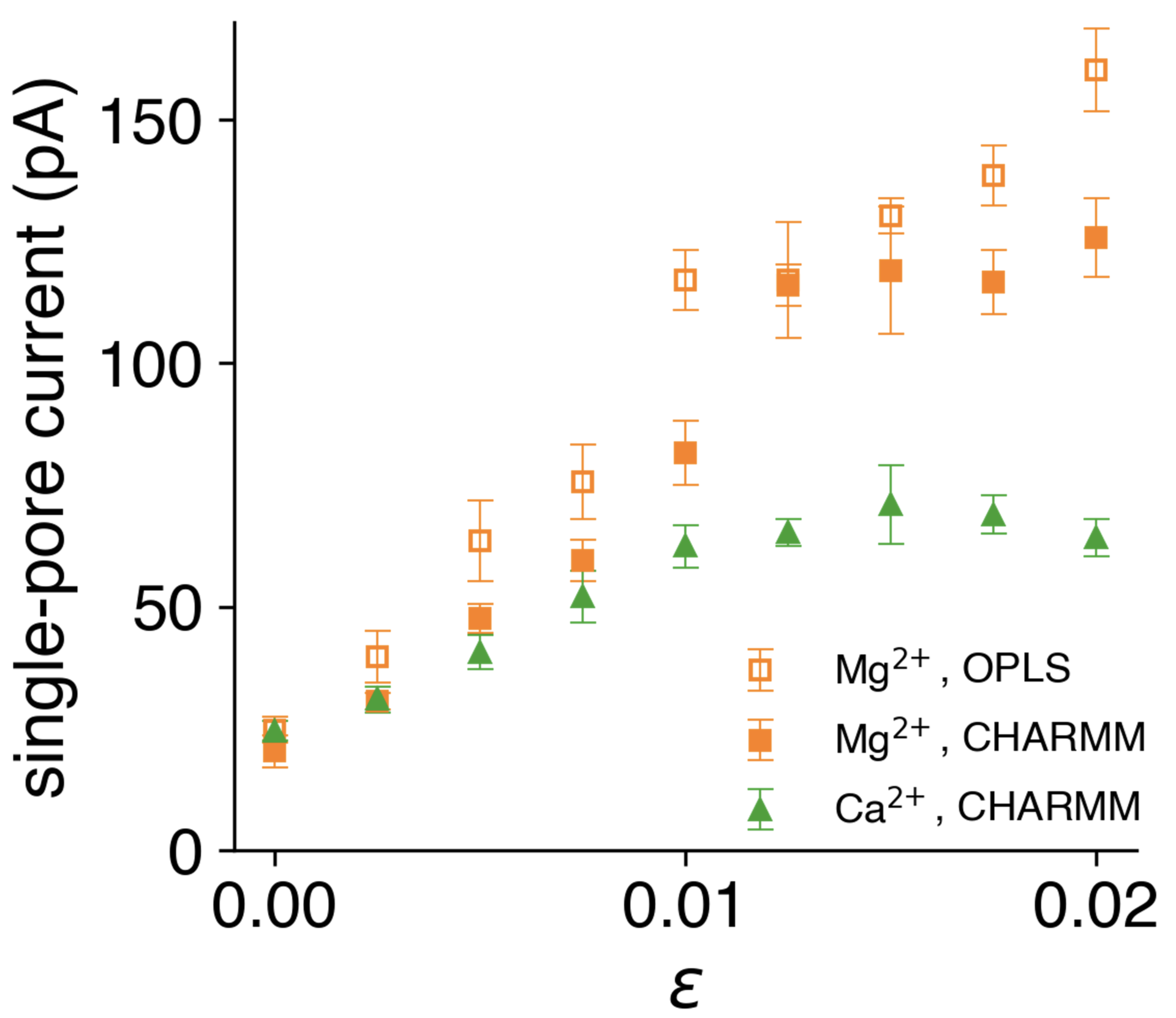}
\caption{Single-pore ion current as a function of isotropic $XY$-strain applied to the hBN membrane, as simulated for various ions and forcefields under a constant electric field $E_Z=0.03$ V/nm. }
\label{fig_s7}
\vspace{18pt} 
\end{figure*}

Interestingly, despite the pore size significantly larger than that in all previous demonstrations of highly mechanosensitive ion transport, divalent cation transport through 16-atom multivacancy pores is exceptionally mechanosensitive. We simulated transport through 4-pore arrays of 16-atom multivacancy pores as described in the main text, except the membrane was larger (to fit a cubic simulation box with 6 nm side length). Transport was initiated by a constant electric field applied in the $Z$-direction. The resulting ionic currents, calculated directly as slopes of the cumulative ion fluxes, are shown in Fig.~\ref{fig_s7}. 

Specifically, at 1\% strain, Mg$^{2+}$ current is shown to increase by a factor of 4.7 (OPLS-AA) and 4.0 (CHARMM 36), relative to the corresponding zero--strain values. Such levels of mechanosensitivity are $\sim2-3$ times higher than those reported earlier for graphene-embedded crown pores~\cite{fang2019highly,sahu2019optimal} and an order of magnitude higher than the previously reported mechanosensitivity for identical pores observed with K$^+$ and Na$^+$~\cite{Noh2025diff2barrier}. Given these levels of transport modulation, if we assume an Arrhenius-type dependence of the ion current on the overall free energy barrier of permeation $\Delta G$, the corresponding barrier reduction is $\delta \Delta G = k_B T \ln(4.7)\approx 1.54k_B T$ and $k_B T \ln(4.0)\approx 1.39k_B T$ for OPLS-AA and CHARMM 36, respectively. Mechanosensitivity is less pronounced for Ca$^{2+}$ ions with an overall less stable FHS and thus a weaker contribution from the enthalpic-entropic competition, with a current ratio of 2.5, corresponding to $\delta \Delta G \approx 0.69k_B T$. Although beyond the scope of this work, strain dependence of the entropic and enthalpic contributions for aqueous Mg$^{2+}$ and Ca$^{2+}$ is straightforward to evaluate in a way similar to that presented in the discussion accompanying Fig. 3 in the main text.

\section{S5. Effects of pore charge and immobilization of water dipoles}
To investigate the impact of pore charge on the entropic contribution to the free energy of Mg$^{2+}$ transport, we conducted a parametric study using 16-atom pores in hBN. We modified the partial charges of the edge nitrogen atoms to maximize the effect by placing the modified charges in close proximity to the ion and its first hydration shell (FHS). With 15 nitrogen atoms at the edge, we adjusted their partial charges to $q_N + \Delta q/15$, where $q_N=-0.6$ (in the units of elementary charge) is the partial atomic charge of edge nitrogen that yields an overall neutral hBN pore structure and $\Delta q$ is the desired overall pore charge. For $\Delta q$ of -2, -1, +1, and +2, each edge nitrogen therefore carried a partial charge of -0.7333, -0.6667, -0.5333, and  -0.4667, respectively. 

\begin{figure*}[ht]
\centering
\includegraphics[width=0.98\textwidth]{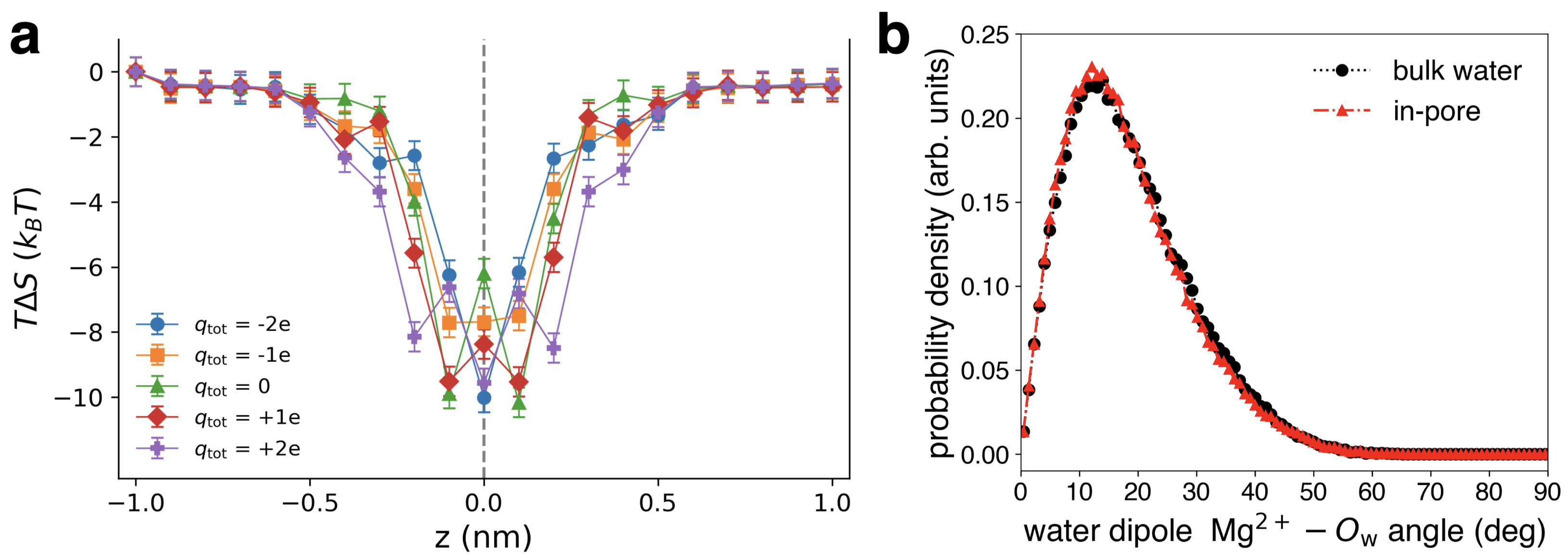}
\caption{Entropic variation for Mg$^{2+}$ permeating the 16-atom pore in hBN for various pore charges (a) and distributions of the angle between the Mg$^{2+}$-O$_w$ vector and the dipole orientation vector for the same water molecule (b). }
\label{fig_s8}
\vspace{18pt} 
\end{figure*}

The resulting entropy estimates are compared in Fig.~\ref{fig_s8} (a). Although the importance of pore charge on the overall value of the barrier (which consists of the enthalpic and entropic contributions) is critically important, the effect specifically on the entropic part is weak for the presented pore charges. As presented, the observed charge-dependent differences are not beyond the uncertainties of our estimates of the primary entropic effect ($k_BT/2$, as stated in the main text).

This observation is not surprising, because the immobilization levels are dictated by general pore confinement, as well as the electrostatic interactions between the FHS and a pore carrying partial charges. The charges are dictated by the chemistry of the material; more specifically, the edge nitrogens carry significant negative partial charge regardless of the total pore charge. This strong non-positivity of the edge nitrogens is a basic property of the material, given the ionic nature of the B-N bond in hBN~\cite{govind2018ab}.

We also investigated whether significant immobilization of water dipoles occurs within the observed FHS immobilization presented above. To do so, we considered the distribution of the angle between the water dipole and the vector formed by the ion and the corresponding water oxygen. The results are shown in Fig.~\ref{fig_s8} (b) and, as presented, the distributions in bulk water and inside the pore are nearly identical. This finding suggests that ion-dipole electrostatics remains the dominating factor in the orientation of water dipoles. Eq. 1 in the main text was used to estimate the corresponding $-T\Delta S_{KL} \approx 0.045k_B T$ \textit{for the entire FHS}, which is two orders of magnitude below the entropic changes arising from the FHS immobilization mechanism presented in this work.

\clearpage

\bibliography{local}